\documentclass{article}

\usepackage{amsmath,bbm,url,fullpage}
\usepackage{stmaryrd}
\usepackage{galois}
\usepackage{subfigure}
\usepackage{multirow}
\usepackage{listings}
\usepackage{graphicx}
\usepackage{algorithm,algorithmic}
\usepackage{pgf,tikz}
\usetikzlibrary{matrix}
\usetikzlibrary{decorations.pathreplacing}
\usetikzlibrary{decorations.pathmorphing}
\usetikzlibrary{backgrounds}
\newcommand\ForAuthors[1]
{\par\smallskip                     
  \begin{center}
    \fbox
    {\parbox{0.9\linewidth}
      {\raggedright\sc--- #1}
    }
  \end{center}
  \par\smallskip                     %
}   

\DeclareSymbolFont{stmry}{U}{stmry}{m}{n}

\newcommand{\Nat}{{\mathbbm N}}
\newcommand{\Real}{{\mathbbm R}}

\newtheorem{definition}{Definition}[section]

\begin{document}

\title{Abstract Fixpoint Computations\\
  with Numerical Acceleration Methods\footnote{The authors want to
    thank Eric Goubault for his helpful discussions and precious
    advices.}}
  
\author{Olivier Bouissou\footnote{CEA, LIST Laboratory for the Modeling and Analysis of Interacting Systems, \texttt{olivier.bouissou@cea.fr}},~ 
  Yassamine Seladji\footnote{CEA, LIST Laboratory for the Modeling and Analysis of Interacting Systems, \texttt{yassamine.seladji@cea.fr}},~
  Alexandre Chapoutot\footnote{Univerist\'e Pierre et Marie Curie -- LIP6, \texttt{alexandre.chapoutot@lip6.fr}}
} 

\date{June 8, 2010}

\maketitle
  
\begin{abstract} 
  Static analysis by abstract interpretation aims at automatically
  proving properties of computer programs. To do this, an
  over-approximation of program semantics, defined as the least
  fixpoint of a system of semantic equations, must be computed. To
  enforce the convergence of this computation, widening operator is
  used but it may lead to coarse results. We propose a new method to
  accelerate the computation of this fixpoint by using standard
  techniques of numerical analysis. Our goal is to automatically and
  dynamically adapt the widening operator in order to maintain
  precision.
  \\

\end{abstract}

\section{Introduction}
\label{sec:introduction}

In the field of static analysis of embedded, numerical programs,
abstract interpretation~\cite{CC77,CC91} is widely used to compute
over-approximations of the set of behaviors of a program. This set is
usually defined as the least fixpoint of a monotone map on an abstract
domain given by the (abstract) semantics of the program. Using
Tarski's theorem~\cite{Tar55}, this fixpoint is computed as the limit
of the iterates of the abstract function starting from the least
element. These iterates build a sequence of abstract elements that
(order theoretically) converge towards the least fixpoint. This
sequence converging often slowly (or even after infinitely many
steps), the theory of abstract interpretation introduces the concept
of \emph{widening}~\cite{CC91}.

A widening operator is a two-arguments function $\nabla$ which tries
to predict the limit of the iterates based on the relative position of
two consecutive iterates. For example, the standard widening operator
on the interval abstract domain consists in comparing the limits of
the intervals and setting the unstable ones to $\infty$ (or
$-\infty$). A widening operator often makes large over-approximation
because it must make the sequence of iterates converge in a finite
time. Over-approximation may be reduced afterward using a
\emph{narrowing} operator but the precision of the final approximation
still strongly depends on the precision of the $\nabla$. Various
techniques have been proposed to improve it. \emph{Delayed} widening
makes use of $\nabla$ after $n$ iteration steps only (where $n$ is a
user-defined integer), thus letting the first loop iterates execute
before trying to predict the limit. Another approach is to use a
widening with \emph{thresholds}~\cite{ASTREE02}: the upper bound of
the interval (for example) is not directly set to $\infty$, but is
successively increased using a set of thresholds that are candidates
for the value of the fixpoint upper bound. In practice, these
techniques are necessary to obtain precise fixpoint approximations for
industrial sized embedded programs. However, they suffer from their
lack of automatization: thresholds must be chosen \emph{a priori} and
are defined by the user (to the best of our knowledge, no methods
exist to automatically find the best thresholds). The delay parameter
$n$ is also to be defined a priori. This makes the use of a static
analyzer difficult as these (non trivial) parameters are often hard to
find.

In this article, we present some ongoing work which shows that it is
possible to use \emph{sequence transformation techniques} in order to
automatically and efficiently derive approximation of the limit of
Kleene iterates. This approximation may not be safe (\textit{i.e.}
may not contain the actual limit), but we show how to use it in the
theory of abstract interpretation.  Sequence transformation techniques
(also known as convergence acceleration methods) are widely studied in
the field of numerical analysis~\cite{BZ91}. They transform a
converging sequence $(x_n)_{n\in\Nat}$ of real numbers into a new
sequence $(y_n)_{n\in\Nat}$ which converges faster to the same limit
(see Section~\ref{sec:acceleration-of-convergence}).  In some cases
(depending on the method), the acceleration is such that
$(y_n)_{n\in\Nat}$ is ultimately constant.  Some recent
work~\cite{brezinski-redivo-08} applied these techniques in the case
of sequences of \emph{vectors} of real numbers: vector sequence
transformations introduce \emph{relations} between elements of the
vectors and perform better than scalar ones. Our main contribution is
to show that we can use these methods in order to improve the fixpoint
computation in static analysis: we define \emph{dynamic} thresholds
for widening that are very close to the actual fixpoint. This
increased precision is obtained because the sequence transformations
use all iterates and \emph{quantitative information} (\textit{i.e.}
relative to the distance between elements) to predict the limit. They
thus have access to more information than the widening operator and
can make better prediction.  In this work, we focus on the interval
domain, but we believe that this work may be applied for any abstract
domain, especially the ones with a \emph{pre-defined shape}
(octagons~\cite{mine:phd}, templates~\cite{SSM05}, etc.).

This article is organized as follows. In
Section~\ref{sec:an-introductive-example}, we explain on a simple
example how acceleration methods may be used to speed-up the fixpoint
computation. In Section~\ref{sec:theoretical-framworks}, we recall the
theoretical basis of this work and present our main theoretical
contribution. Section~\ref{sec:experimentations} presents some early
experiments on various floating-point programs that show the interest
of our approach, while Sections~\ref{sec:related-work}
and~\ref{sec:conclusion} discuss related works and perspectives.

\noindent {\bf Notations.} In the rest of this article, $(x_n)$ will
denote a sequence of real numbers (\textit{i.e.}
$(x_n)\in\Real^\Nat$), while $(\boldsymbol{x}_n)$ denotes a sequence
of vector of real numbers (\textit{i.e.}
$(\boldsymbol{x}_n)\in\bigl(\Real^p\bigr)^\Nat$ for some
$p\in\Nat$). The symbol $X_n$ will be used to design \emph{abstract
  iterates}, i.e. $X_n\in A$ for some abstract lattice $A$.

\section{An introductive example}
\label{sec:an-introductive-example}

In this section, we explain, using a simple example, how sequence
acceleration techniques can be used in the context of static
analysis. In short, our method works as follows: let $(X_n)$ be a
sequence of intervals computed by the Kleene iteration and that is
chosen to be widened (see~\cite{Bour93} for details on how to chose
the widening points). We extract from $(X_n)$ a vector sequence
$(\boldsymbol{x}_n)$: at stage $k$, $\boldsymbol{x}_k$ is a vector
that contains the infimum and supremum of each variable of the
program. As Kleene iterates converge towards the least fixpoint of the
abstract transfer function, the sequence $(\boldsymbol{x}_n)$
converges towards a limit $\boldsymbol{x}$ which is the vector
containing the infimum and supremum of this fixpoint. We then compute
an accelerated sequence $(\boldsymbol{y}_n)$ that converges towards
$\boldsymbol{x}$ faster than $(\boldsymbol{x}_n)$. Once this sequence
has reached its limit (or is sufficiently close to it), we use
$\boldsymbol{x}$ as a threshold for a widening on $(\boldsymbol{x}_n)$
and thus obtain, in a few steps, the least fixpoint. In the rest of
this section, we detail these steps.

\null
\noindent {\bf The program.} We consider a linear program which
iterates the function $F(X)=A\cdot X+B\cdot U$ where $A$, $B$ and $U$
are constant matrices and $X$ is the vector of variables (see
Figure~\ref{fig:prog-intro}). Initially, we have
$\texttt{x1}\in[1,2],\ \texttt{x2}\in[1,4],\ \texttt{x3}\in[1,20],\
\texttt{u1}\in[1,6],\ \texttt{u2}\in[1,4]$ and $\texttt{u3}\in[1,2]$.
Using an existing analyzer working on the interval abstract domain, we
showed that this program converges in 55 iterations (without widening)
and obtained the invariant $[-5.1975,8.8733]$ for \verb+x1+ at line
$2$.

\begin{figure}[tbp]
  \centering
 \begin{minipage}{0.9\linewidth}
   \begin{lstlisting}[numbers=left,frame=single]
 while (1) {
   xn1 = -0.4375 * x1+ 0.0625 * x2 + 0.2652 * x3 + 0.1 * u1;
   xn2 = 0.0625 * x1 + 0.4375 * x2 + 0.2652 * x3 + 0.1 * u2;
   xn3 = -0.2652 * x1 + 0.2652 * x2 + 0.375 * x3 + 0.1 * u3;
   x1 = xn1; x2 = xn2; x3 = xn3;
  }
\end{lstlisting}  
  \end{minipage}
  \caption{A simple linear program.}
  \label{fig:prog-intro}
\end{figure}

\begin{figure}[btp]
  \centering
  \begin{minipage}{0.48\linewidth}
    \includegraphics[scale=0.25,angle=-90]{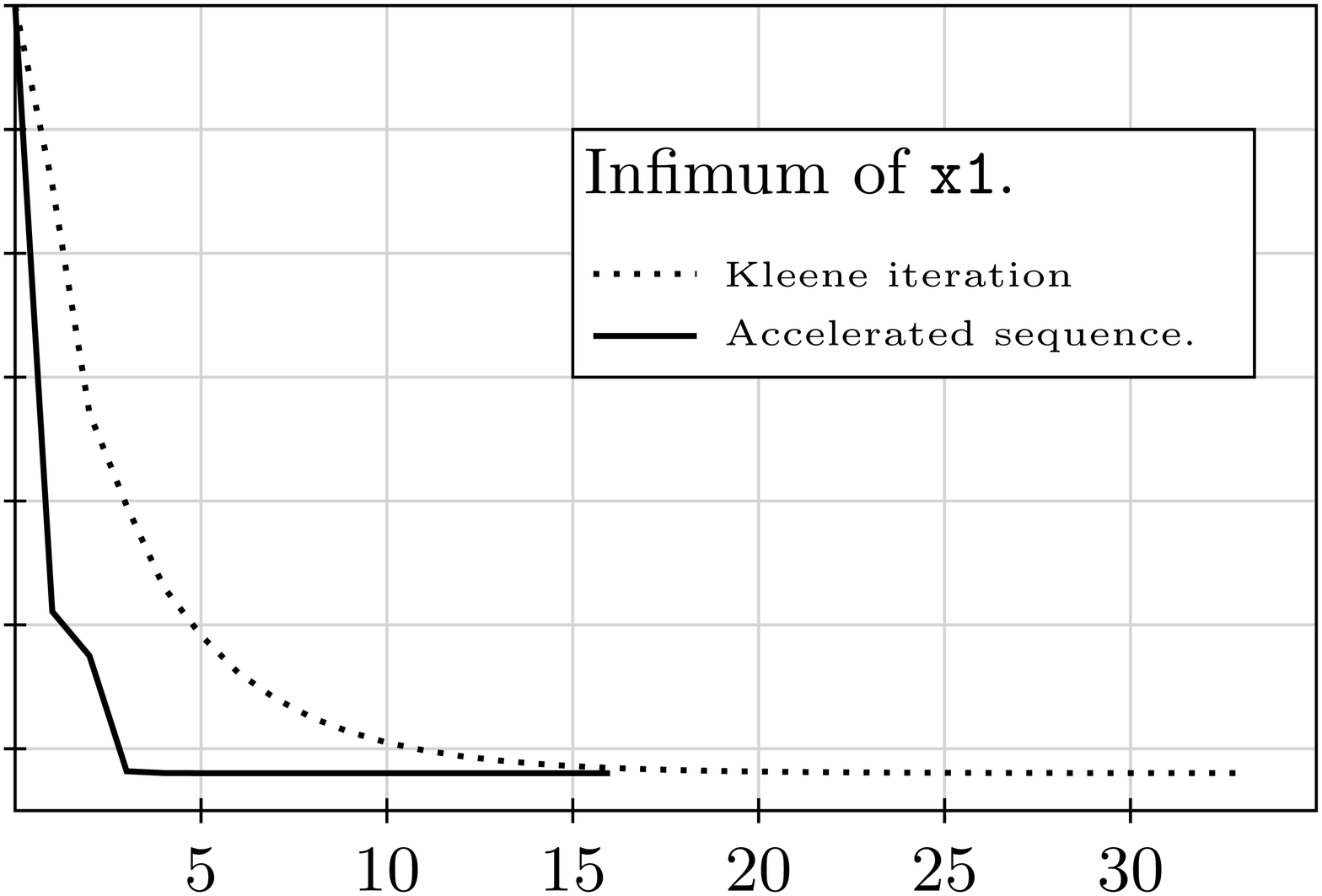}
  \end{minipage}
  \begin{minipage}{0.48\linewidth}
    \includegraphics[scale=0.25,angle=-90]{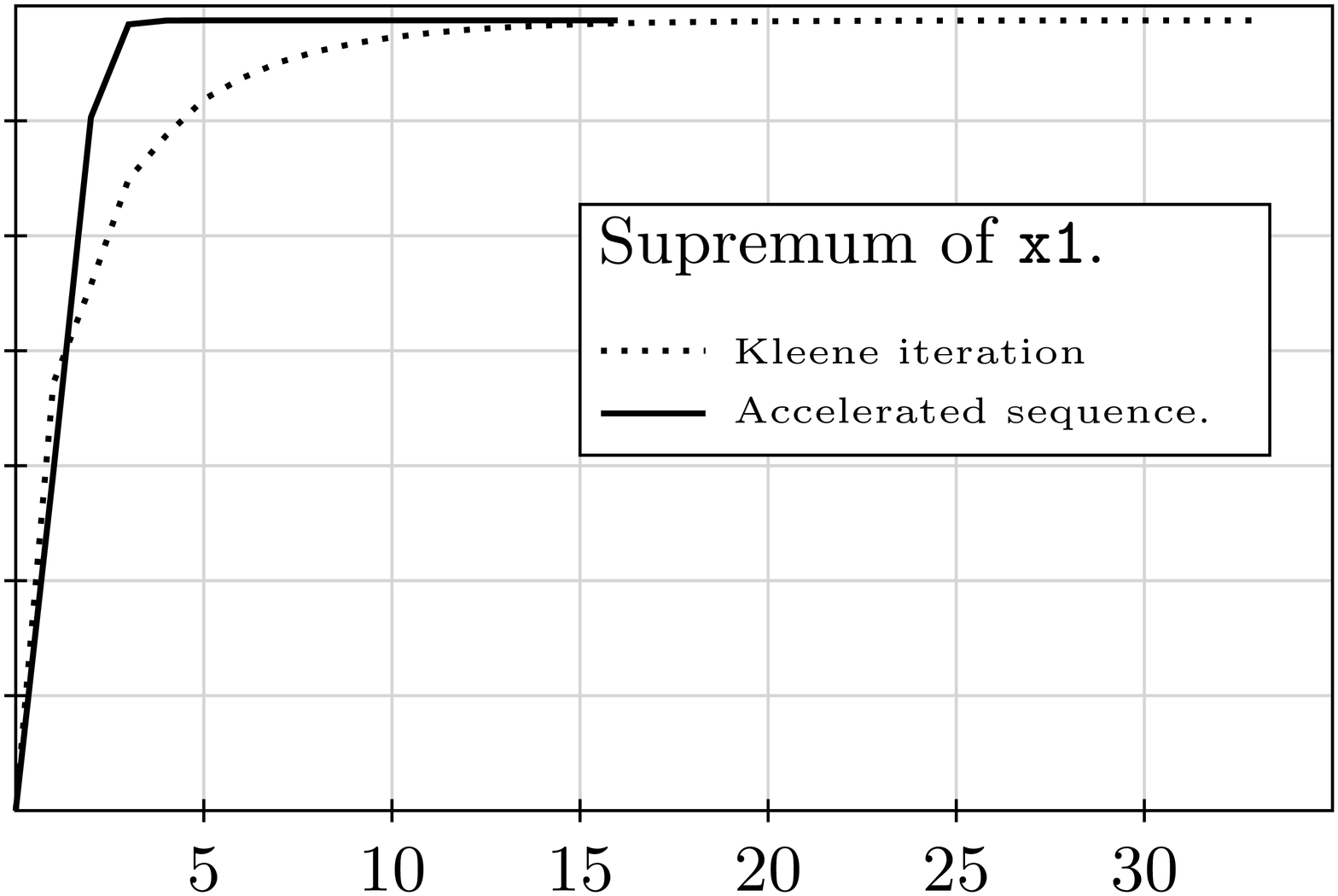}
  \end{minipage}
  \caption{Sequences extracted from the program of
    Figure~\ref{fig:prog-intro} and their accelerated version.}
  \label{fig:exemple-intro}
\end{figure}

\null
\noindent {\bf Extracting the sequences.} From this program, we can
define a vector sequence of size $6$,\\
$\boldsymbol{x}_n=\bigl(
\underline{x^1_n},\overline{x^1_n},
\underline{x^2_n},\overline{x^2_n}, \underline{x^3_n},\overline{x^3_n}
\bigr) $, which represent the evolution of the suprema and infima of
the variables \verb+x1+, \verb+x2+ and \verb+x3+ at line $2$. For
example, the sequence $(\overline{x^1_n})$ is recursively defined by:

{\small\begin{equation}
\overline{x^1_{n+1}}=\max\Bigl(\ 
\overline{x^1_n}\ 
,\ 
-0.4375*\underline{x^1_n} + 0.0625*\overline{x^2_n}+
0.2652*\overline{x^3_n} + 0.1*\overline{u_1}\ 
\Bigr)\;.
\label{eq:x1n}
\end{equation}
}%
Note that we are not interested in the formal definition of these
sequences (as given by Equation~\eqref{eq:x1n}), but only in their
numerical values that are easily extracted from Kleene
iterates. Each sequence
$(\overline{x^i_n})$ (resp.~$(\underline{x^i_n})$) is increasing
(resp.~decreasing) and the sequence $(\boldsymbol{x}_n)$ converge
towards a vector $\boldsymbol{x}$ containing the infimum and supremum of
the fixpoint (see Figure~\ref{fig:exemple-intro}, dotted lines).

\null
\noindent {\bf Accelerating the sequences.} We then used the
\emph{vector $\varepsilon$-algorithm}~\cite{brezinski-redivo-08} to
build a new sequence that converges faster towards $\boldsymbol{x}$.
This method works as follows (a more formal definition will be given
in Section~\ref{sec:acceleration-of-convergence}): it computes a
series of sequences $(\boldsymbol{\varepsilon}^k_n)$ for $k=1,2,\dots$
such that each sequence $(\boldsymbol{\varepsilon}^k_n)$ for $k$
\emph{even} converges towards $\boldsymbol{s}$ and the \emph{diagonal}
$(\boldsymbol{d}_n)=(\boldsymbol{\varepsilon}^n_0)$ also converges
towards $\boldsymbol{s}$. This diagonal sequence is the result of the
$\varepsilon$-algorithm and is called the \emph{accelerated
  sequence}. It converges faster than the original sequence: in only
$8$ iterates, it reached the fixpoint and stayed constant (see
Figure~\ref{fig:exemple-intro}, bold lines).

\begin{figure}[tbp]\centering
  \vspace{-2.5cm}
  \includegraphics[scale=0.5,angle=-90]{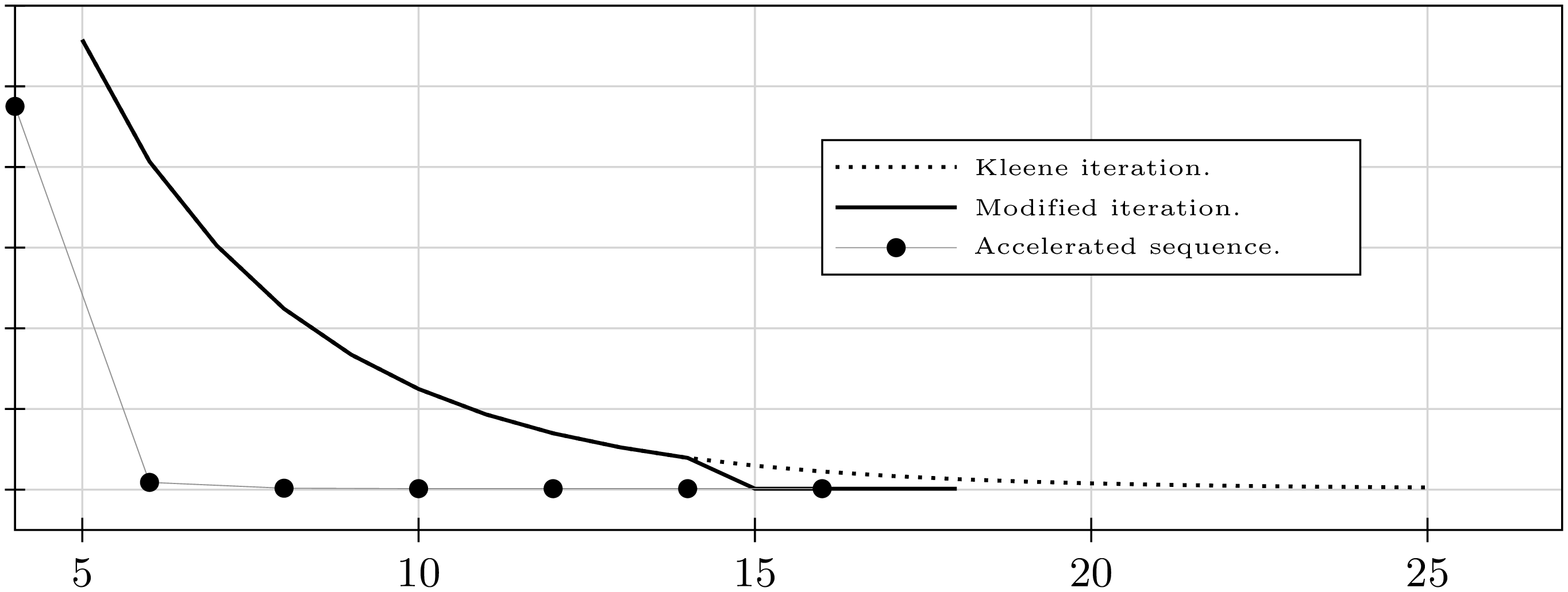}
  \vspace{-2.5cm}
     \caption{Infimum value of $x_1$. We only display the iterates $5$
       to $25$. At the $15$th iteration, the accelerated value is used
       as a widening with thresholds, and the iteration stops after
       $18$ steps.}
  \label{fig:exemple-intro-sumup}
\end{figure}

\null
\noindent {\bf Using the accelerated sequence.}
When the accelerated sequence reaches the limit (or is sufficiently
close to it), we modify the Kleene iteration and directly jump to the
limit. Formally, if the limit is $(\underline{x_1},\overline{x_1},
\underline{x_2},\overline{x_2}, \underline{x_3},\overline{x_3})$ and
if the current Kleene iterate is $X_p$, we construct the abstract
element $X$ whose bounds are $\underline{x_1},\ \overline{x_1},\dots$
and set $X_{p+1}=X_p\cup X$ and re-start Kleene iteration from
$X_{p+1}$. In this way, we remain sound ($X_p\subseteq X_{p+1}$) and
we are very close to the fixpoint, as $X\subseteq X_{p+1}$. In this
example, Kleene iteration stopped after $2$ steps and reached the same
fixpoint as the one obtained without
widening and acceleration. Figure~\ref{fig:exemple-intro-sumup} shows
the original 
Kleene iteration and the modified one, for the infimum of variable
\verb+x1+. Let us recall that the Kleene iteration needed $55$ steps
to converge, where the modified iteration stops after $18$ steps.

\section{Theoretical frameworks}
\label{sec:theoretical-framworks}

In this section, we briefly recall the basics of abstract
interpretation, with an emphasis on the widening operator. Next we
present in more details the theory of sequence
transformations. Finally, we give our main contribution showing how
sequence transformations are used in abstract interpretation theory.

\subsection{Overview of the abstract interpretation theory}
\label{sec:abstract-interpretation-theory}

Abstract interpretation is a general method to compute
over-approximations of program semantics where the two key ideas are:
\begin{itemize}
\item \textit{Safe abstractions} of sets of states thanks to Galois
  connections. More precisely let $\langle C, \sqsubseteq_C \rangle$
  be the lattice of concrete states and let $\langle A, \sqsubseteq_A
  \rangle$ be the lattice of abstract states. $A$ is a safe
  abstraction of $C$ if there exists a Galois connexion $\langle C,
  \sqsubseteq_C \rangle \galois{\gamma}{\alpha} \langle A,
  \sqsubseteq_A \rangle$, \textit{i.e.} there exist monotone maps
  $\alpha$ and $\gamma$ such that $\forall c \in C, \forall a \in A,
  \alpha (c) \sqsubseteq_A a \Leftrightarrow c \sqsubseteq_C
  \gamma(a)$.
\item An \textit{effective} computation method of the abstract
  semantics with, in general, a widening operator. The semantics of a
  program is defined as the smallest solution of a recursive system of
  semantic equations $F$. Hence, the abstract program semantics is a
  set of states $X$ of a lattice $\langle A, \sqsubseteq_A\rangle$
  such that $X = F(X)$ where $F$ is monotone. The solution $X$ is
  iteratively constructed by $X_{i+1} = X_i \sqcup F(X_i)$, starting
  from $X_0 = \bot$. The value $\bot$ denotes the smallest element of
  $A$ and the operation $\sqcup$ denotes the join operation of
  $A$. The sequence $(X_n)$ defines an increasing chain of elements of
  $A$. This chain may be infinite, so to enforce the convergence of
  this sequence, we usually substitute the operator $\sqcup$ by a
  \textit{widening operator} $\nabla$, see
  Definition~\ref{def:widening}, that is an over-approximation of
  $\sqcup$.
\end{itemize}

\begin{definition}[Widening operator \cite{CC77}]
  \label{def:widening}
  Let $\langle A, \sqsubseteq_A \rangle$ be a lattice. The map
  $\nabla: A\times A\rightarrow A$ is a widening operator iff
  \textit{i)}~$\forall v_1, v_2 \in A$, $v_1 \sqcup v_2 \sqsubseteq_A
  v_1 \nabla v_2$. \textit{ii)}~For each increasing chain $v_0
  \sqsubseteq_A \cdots \sqsubseteq_A v_n \sqsubseteq_A \cdots$ of $A$,
  the increasing chain defined by $s_0 = v_0$ and $s_n = s_{n-1}
  \nabla v_n$ is stationary: $\exists n_0, \forall n_1, n_2, (n_2 >
  n_1 > n_0) \Rightarrow s_{n_1} = s_{n_2}$.
\end{definition}

The widening operator plays an important role in static analysis
because, thanks to it, we are able to consider infinite state spaces. As
a consequence, many abstract domains are associated with a widening
operator.  For example the classical widening of the interval domain
is defined by:

{\small
  \begin{displaymath} [a, b] \nabla [c, d] = \left[
      \begin{cases}
        a & \text{if } a \le c
        \\
        -\infty & \text{otherwise}
      \end{cases},\quad
      \begin{cases}
        b & \text{if } b \ge d
        \\
        +\infty & \text{otherwise}
      \end{cases}
    \right]\enspace.
  \end{displaymath}
}%
Note that we only consider two consecutive elements to extrapolate the
potential fixpoint. The main drawback with this widening is that it may
generate too coarse results by going quickly to infinity. A solution of
this is to add intermediate steps among a finite set $T$; that is the
idea behind the \textit{widening with 
  thresholds} $\nabla_T$. For the interval domain, it is defined
\cite{BCC+03} by: 

{\small
  \begin{displaymath} [a, b] \nabla_T [c,d] = \left[
      \begin{cases}
        a & \text{if } a \le c
        \\
        \max \{ t \in T: t \le c \} & \text{otherwise}
      \end{cases}
      ,\quad
      \begin{cases}
        b & \text{if } b \ge d
        \\
        \min \{ t \in T: t \ge d\} & \text{otherwise}
      \end{cases}
    \right]\enspace.
  \end{displaymath}
}%
While widening with thresholds gives better results, we are facing
with the problem to define \textit{a priori} the set $T$. Finding
relevant values for $T$ is a difficult task for which, to the best
of our knowledge, no automatic solution exists.


\subsection{Acceleration of convergence}
\label{sec:acceleration-of-convergence}

We give an overview of the techniques of acceleration of convergence
in numerical analysis~\cite{BZ91}. The goal of convergence
acceleration techniques, also named \textit{sequence transformations},
is to increase the rate of convergence of a sequence. Formally, let
$\bigl(D,d\bigr)$ be a metric space, \textit{i.e.} a set $D$ with a
distance $d:D\rightarrow\Real^+$ ($D$ will be $\Real$ or $\Real^p$ for
some $p\in\Nat$). The set of sequences over $D$ (denoted $D^\Nat$) is
the set of functions between $\Nat$ and $D$. A sequence $(x_n)\in
D^\Nat$ converges to $\ell$ iff we have
$\lim_{n\rightarrow\infty}d(x_n,\ell)=0$. A \emph{sequence
  transformation} is a function $T:D^\Nat\rightarrow D^\Nat$ ($T$
designs a particular acceleration method) such that whenever $(x_n)$
converges to $\ell$ then $(y_n)=T(x_n)$ also converges to $\ell$ and
$\lim_{n\rightarrow\infty}\frac{d(y_n,\ell)}{d(x_n,\ell)}=0$. This
means that $(y_n)$ is asymptotically closer to $\ell$ than $(x_n)$. An
important notion for a sequence transformation $T$ is its kernel $K_T$
which is the set of sequences $(x_n)$ for which $T(x_n)$ is ultimately
constant. We now present some acceleration methods that we used in our
experimentation. For more details, we refer to~\cite{BZ91}.

\null
\noindent {\bf The Aitken $\Delta^2$-method.} 
It is probably the most famous sequence transformation. Given a
sequence $(x_n)\in\Real^\Nat$, the accelerated sequence $(y_n)$ is
defined by: $\forall n\in\Nat,\ y_n = x_n -
\frac{x_{n+1}-x_n}{x_{n+2}-2x_{n+1}+x_n}$.  It should be noted that in
order to compute $y_n$ for some $n\in\Nat$, three values of $(x_n)$
are required: $x_n$, $x_{n+1}$ and $x_{n+2}$. The kernel
$K_{\Delta^2}$ of this method is the set of all sequences of the form
$x_n=s+a.\lambda^n$ where $s$, $a$ and $\lambda$ are real constants
such that $a\neq 0$ and $\lambda\neq 1$ (see~\cite{Brez07}). The
Aitken $\Delta^2$-method is an efficient method for accelerating
sequences, but it highly suffers from numerical instability when
$x_n$, $x_{n+1}$ and $x_{n+2}$ are close to each other.

\begin{figure}[hb]
  \begin{minipage}{0.6\linewidth}
    \includegraphics[scale=0.25,angle=-90]{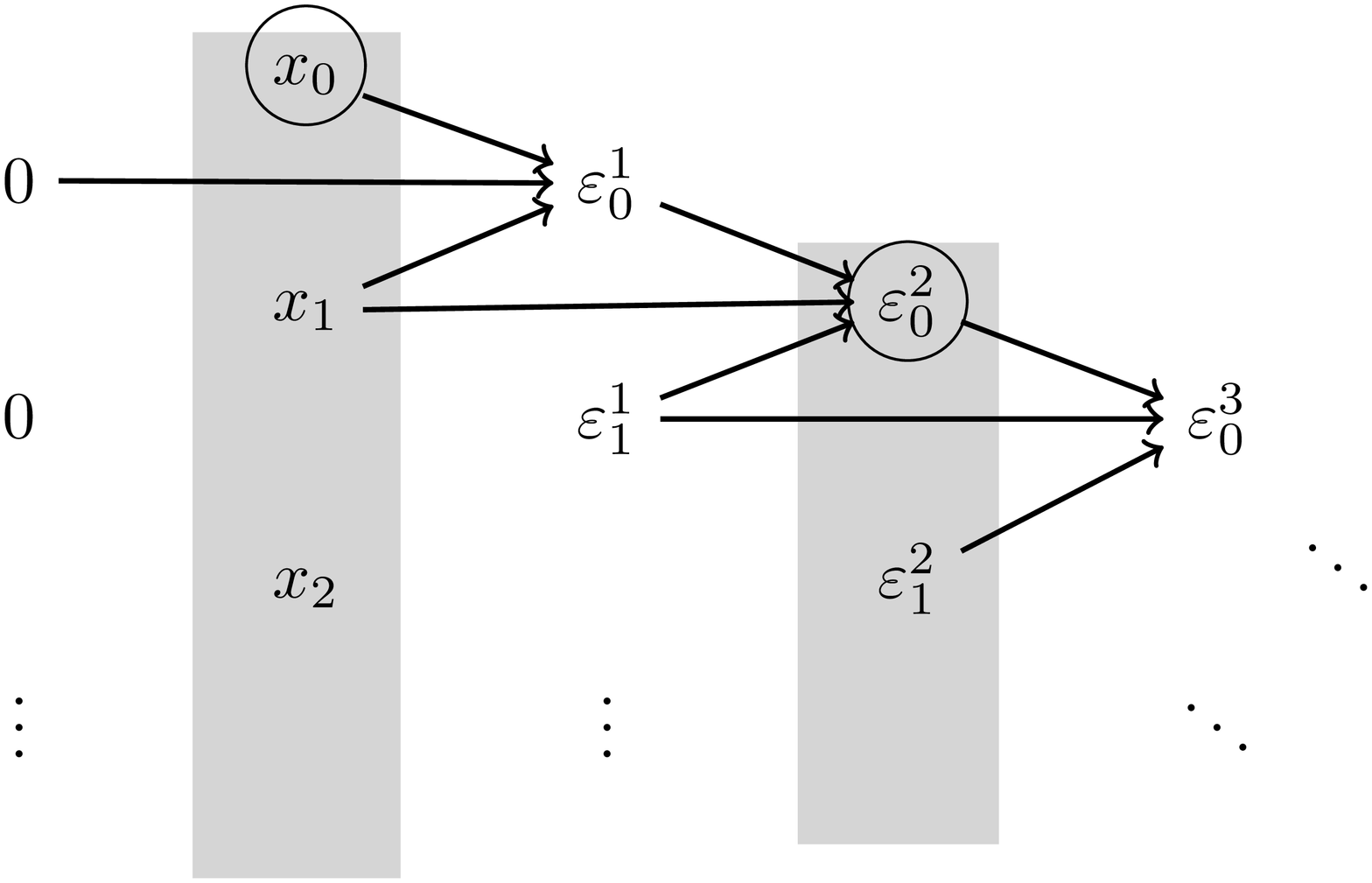}
  \end{minipage}
  \begin{minipage}{0.39\linewidth}
    {\small Arrows depict dependencies: the element at the beginning of
      the arrow is required to compute the element at the end.  For
      example,
      \begin{align*}
        \varepsilon_0^2 & = \varepsilon_1^0 +
        \frac{1}{\varepsilon_1^1-\varepsilon_0^1}
        \\
        & = x_1 + \frac{1}{\varepsilon_2^{-1} 
          + \frac{1}{\varepsilon_2^0-\varepsilon_1^{0}} 
          - \varepsilon_1^{-1} + \frac{1}{\varepsilon_1^0 - \varepsilon_0^0}}
        \\
        & = x_1 + \frac{1}{\frac{1}{x_2-x_1} - \frac{1}{x_1 - x_0} }
      \end{align*}
    }
  \end{minipage}
  \caption{The $\varepsilon$-table}
  \label{fig:epsilon-table}
\end{figure}

\null
\noindent {\bf The $\varepsilon$-algorithm.}
It is often cited as the best general purpose sequence transformation
for slowly converging sequences \cite{Wynn61}. From a converging
sequence $(x_n)\in\Real^\Nat$ with limit $\ell$, the
$\varepsilon$-algorithm builds the following sequences:

{\small
  {\begin{align} (\varepsilon^{-1}_n) &: \forall n\in\Nat,
      \varepsilon^{-1}_n=0,\label{eq:epsilon-1}
      \\
      (\varepsilon^{0}_n) &: \forall n\in\Nat,
      \varepsilon^{0}_n=x_n,\label{eq:epsilon-2}
      \\
      (\varepsilon^k_n) &: \forall k\geq 1,\ n\in\Nat,\ \varepsilon^{k+1}_n=
      \varepsilon^{k-1}_{n+1}+\bigl(\varepsilon^k_{n+1}-\varepsilon^k_n\bigr)^{-1}
      \label{eq:epsilon-3}
  \end{align}
}}%
The sequence $(\varepsilon^k_n)$ is called the $k$-th column, and its
construction can be graphically represented as on
Figure~\ref{fig:epsilon-table}. The \emph{even} columns
$(\varepsilon^{2k}_n)$ (in gray on Figure~\ref{fig:epsilon-table})
converge faster to $\ell$. The \emph{even} diagonals
$(\varepsilon^{2k}_{n})_{k\in\Nat})$ also converge faster to $\ell$. In
particular, the first diagonal (circled on
Figure~\ref{fig:epsilon-table}) converges very quickly to $\ell$, and
it is the accelerated sequence. Let us remark that in order to compute
the $n$-th element of that sequence, $2n$ elements of $(x_n)$ are
required.

\null
\noindent {\bf Acceleration of vector sequences.}
Many acceleration methods were designed to handle scalar sequences of
real numbers. For almost each of these methods, extensions have been
proposed to handle vector sequences (see~\cite{GM92} for a review of
them). The simplest, yet one of the most powerful, of these methods is
the \emph{vector $\varepsilon$-algorithm} (VEA). Given a vector sequence
$(\boldsymbol{x}_n)$, the VEA computes a series of vector sequences
$(\boldsymbol{\varepsilon}^k_n)$ using
Equations~\eqref{eq:epsilon-1}-\eqref{eq:epsilon-3} where the
arithmetic operations $+$ and $-$ are computed component-wise and the
inverse of a vector $\boldsymbol{v}$ is computed as
$\boldsymbol{v}^{-1}=\boldsymbol{v}/(\boldsymbol{v}\cdot\boldsymbol{v})$,
with $/$ being the component-wise division and $\cdot$ the scalar
product. The VEA differs from a component-wise
application of the (scalar) $\varepsilon$-algorithm as it introduces
\emph{relations} between the components of the vector: the scalar
product $\boldsymbol{v}\cdot\boldsymbol{v}$ computes a global
information on the vector $\boldsymbol{v}$ which is
propagated to all components. Our experiments show that this
algorithm works better than a component-wise application of the
$\varepsilon$-algorithm. The kernel $K_\varepsilon$ of the VEA contains all
sequences of the form $\boldsymbol{x}_{n+1}=A\boldsymbol{x}_n+B$,
where $A$ is a constant matrix and $B$ a constant
vector~\cite{brezinski-redivo-08}.

\subsection{Our contribution}
\label{sec:our-contribution}

In this section, we combine acceleration methods with the abstract fixpoint
computation. Our goal is to be as non-intrusive as
possible in the classical iterative scheme. In this way, our method
can be implemented with minor adaptations in current static analyzers.

\null
\noindent {\bf Methodology.}
As seen in Section~\ref{sec:abstract-interpretation-theory}, the
Kleene iteration for finding the least fixpoint computes with abstract
values from some abstract lattice $A$. In order to use acceleration
techniques on the abstract iterates, we need to extract from the
abstract elements $X_n\in A$ a vector of real numbers. Thus, we obtain
a sequence of real vectors that we can accelerate, and we quickly
reach its limit. We then construct an abstract element $X$ that
corresponds to this limit and use it as a candidate for the least
fixpoint. This process of transforming an abstract value into a real
vector and back is formalized by the notion of \emph{extraction} and
\emph{combination} functions that are given in
Definition~\ref{def:extraction-combination}.

\begin{definition}[Extraction and combination.]
  \label{def:extraction-combination}
  Let $\langle A,\sqsubseteq_A \rangle$ be an abstract domain, and let
  $p\in\Nat$. The functions $\Lambda_A:A\rightarrow\Real^p$ and
  $\Upsilon_A:\Real^p\rightarrow A$ are called extraction and
  combination function, respectively, iff for each sequence $X_n\in
  A^\Nat$ that \emph{order theoretically converges}, \emph{i.e.}
  $\sqcup_{n\in\Nat}X_n=X$ for some $X\in A$, then the sequence
  $\Lambda_A(X_n)\in\bigl(\Real^p\bigr)^\Nat$ \emph{converges for the
    usual metric on $\Real^p$}, \emph{i.e.}
  $\lim_{n\rightarrow\infty}\Lambda_A(X_n)=S$, and
  $X\sqsubseteq_A\Upsilon_A(S)$.
\end{definition}

Intuitively, these functions transpose the convergence of the sequence
of iterates into the theory of real sequences, in such a way that the
real sequence does not lose any information. Note that the order on
$\Real^p$ induced by the usual metric is unrelated with the order
$\sqsubseteq_A$ on $A$, so the notion of extraction and combination is
different from the notion of Galois connection used to compare
abstract domains. For the interval domain $I=\mathbbm{I}^v$, where $v$
is the number of variables of the program and $\mathbbm{I}$ is the set
of floating-point intervals, the extraction and the combination
functions are defined in Equation.~\eqref{eq:extraction-combination}.

For other domains, these functions must be designed specifically. For
example, we believe that such functions can be easily defined for the
octagon abstract domain~\cite{mine:phd}: the function $\Lambda$
associates with a \emph{difference bound matrix} a vector containing
all its coefficients. Special care should be taken in the case of
infinite coefficients. More generally, we believe that for domains
with a pre-defined shape, the functions $\Lambda$ and $\Upsilon$ can
be easily defined. Note that if there is a Galois connection
$(\alpha_I,\gamma_I)$ between a domain $A$ and the interval domain
$I$, the extraction and combination functions can be defined as
$\Lambda_A=\Lambda_I \circ \alpha_I$ and
$\Upsilon_A=\gamma_I\circ\Upsilon_I$.
We use this method in the last experiment in
Section~\ref{sec:butterworth-order-2}.

{\small
  \begin{equation}
    \label{eq:extraction-combination}
    \begin{aligned}
      \Lambda_I & : \left\{
        \begin{array}{rcl}
          I & \rightarrow & \Real^{2v}
          \\
          (i_1,\dots,i_v)
          &
          \mapsto
          &
          \bigl(\overline{i_1},\underline{i_1},
          \dots,\overline{i_V},\underline{i_V}\bigr)
        \end{array}
      \right.
      \\
      \Upsilon_I & : \left\{
        \begin{array}{rcl}
          \Real^{2v}
          &
          \rightarrow
          &
          I
          \\
          (x_1,x_2,\dots,x_{2v-1},x_{2v})
          &
          \mapsto
          &
          \bigl([x_1,x_2],\dots,[x_{2v-1},x_{2v}]\bigr)
        \end{array}
      \right.
    \end{aligned}
  \end{equation}
}

\null
\noindent {\bf Accelerated abstract fixpoint computation.}
We describe the insertion of acceleration methods in the Kleene
iteration process in
Algorithm~\ref{algo:accelerated-abstract-fixpoint-computation}. We
compute in parallel the sequence $(X_n)$ coming from the Kleene's
iteration and the accelerated sequence $(\boldsymbol{y}_n)$ computed from an
accelerated method. Once the sequence $(\boldsymbol{y}_n)$ seems to
converge, that 
is the distance between two consecutive elements of $(\boldsymbol{y}_n)$
is smaller 
than a given value $\delta$, we combine the two sequences. That is we
compute the upper bound of the two elements of the current
iteration. Note that the monotonicity of the computed sequence $(X_n)$
is still guaranteed.

\begin{algorithm}
  \caption{Accelerated abstract fixpoint computation}
  \label{algo:accelerated-abstract-fixpoint-computation}
  \begin{algorithmic}[1]
    \REPEAT
    \STATE{$X_i := X_{i-1}\ \sqcup\ F(X_{i-1})$}
    \STATE{$\boldsymbol{y}_i := \text{Accelerate} 
      \left(\Lambda_A(X_0), \dots, \Lambda_A(X_i)\right)$}
    \IF{$|| \boldsymbol{y}_i - \boldsymbol{y}_{i-1} ||\ \le\ \delta$}
    \STATE{$X_i := X_i\ \sqcup\ \Upsilon_A(\boldsymbol{y}_i)$}
    \ENDIF
    \UNTIL{$X_i \sqsubseteq X_{i-1}$}
  \end{algorithmic}
\end{algorithm}

The use of acceleration methods may be seen as an automatic delayed
application of the widening with thresholds. Let us remark that we are
not guaranteed to terminate in finitely many iterations: we know that
asymptotically, the sequence $\boldsymbol{y}_i$ from
Algorithm~\ref{algo:accelerated-abstract-fixpoint-computation} gets
closer and closer to the fixpoint, but we are not guaranteed that it
reaches it. To guarantee termination of the fixpoint computation, we
have to use more ``radical'' widening thresholds, for example after
$n$ applications of the accelerated method. So this method cannot be a
substitute to widening, but it improves it by reducing the number of
parameters (delay and thresholds) that a user must define.

\section{Experimentation}
\label{sec:experimentations}

To illustrate our acceleration methods, we used a simple static
analyzer\footnote{This analyzer is based on Newspeak,
  \url{http://penjili.org/newspeak.html}{http://penjili.org/newspeak.html},
  the authors thank especially Sarah Zennou for her technical help.}
working on the interval abstract domain that handles C programs
without pointers and associated it with our OCaml library of
acceleration methods that transform an input sequence (given as a
sequence of values) into its accelerated version. The obtained results
are presented in the following sections.

\subsection{Butterworth order~$1$}
\begin{figure}[btp]
  \centering
  \begin{minipage}{0.48\linewidth}
\begin{lstlisting}[frame=single]
  x1 = 0; y = 0; xn1 = 1;
  for (i=0;i<200;i++) {
    /*!npk u between 1 and 2 */
    xn1 = 0.90480*x1 + 0.95240*u;
    y = 0.09524*x1 + 0.04762*u;
    x1 = xn1;
  } 
\end{lstlisting}
  \end{minipage}
\hfill
\begin{minipage}{0.48\linewidth}
\includegraphics[scale=0.25,angle=-90]{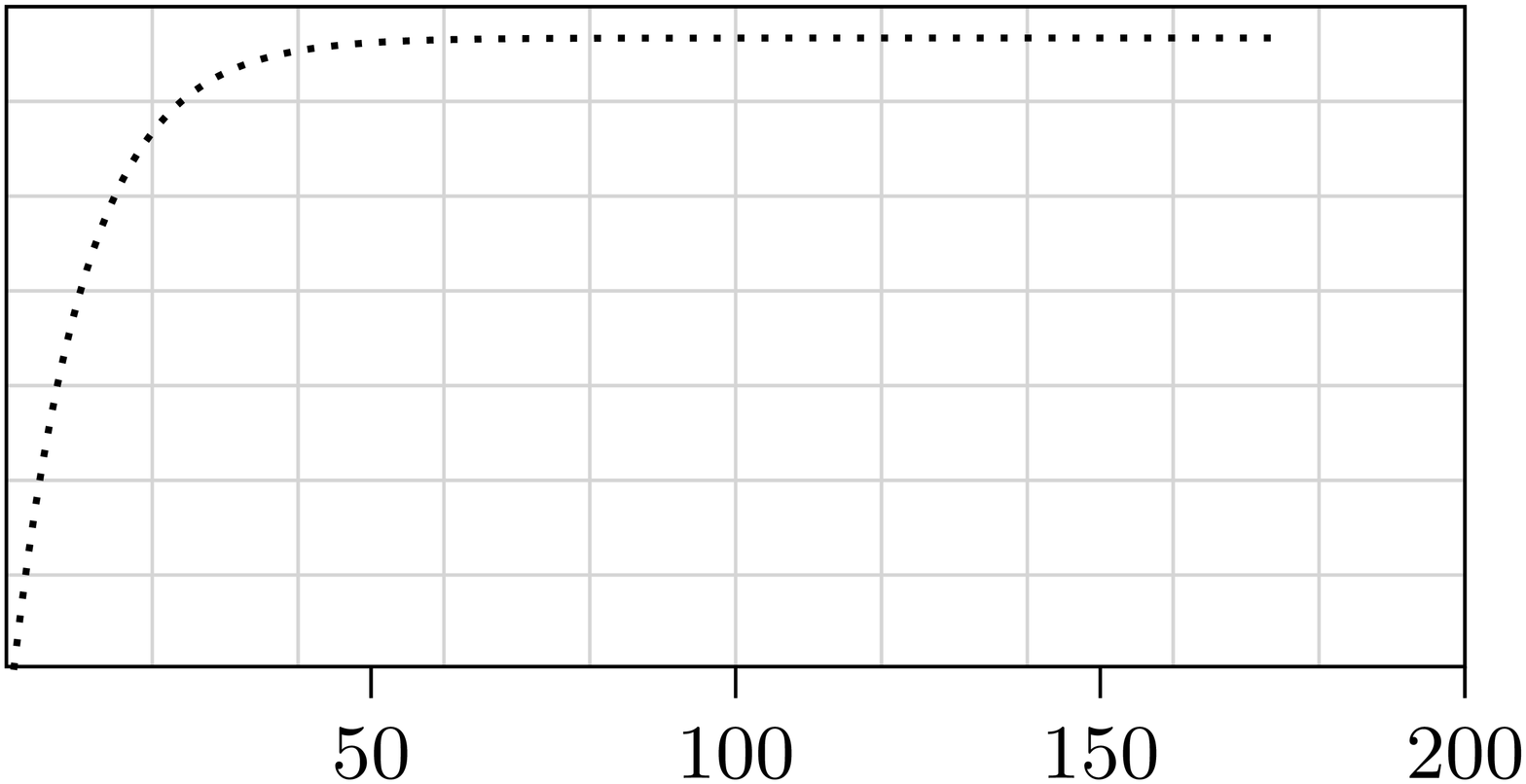}
\end{minipage}
    \caption{The Butterworth program (left) and the sequence of
      supremum of variable \texttt{x1} (right).}
  \label{fig:prog-butterworth}
\end{figure}

To test the acceleration method, we use a first-order Butterworth
filter (see Figure~\ref{fig:prog-butterworth}, left). This filter is
designed to have a frequency response which is as flat as
mathematically possible in the band-pass and is often used in embedded
systems to treat the input signals for a better stability of the
program.

The static analysis of this program using the interval abstract domain
defines $10$ sequences, two for each variable (\verb,x1,, \verb,xn1,,
\verb,y,, \verb,u,, \verb,i,). These sequences converge toward the
smallest fixpoint after a lot of iterations, our acceleration methods
allow to obtain the same fixpoint faster. In this example, we
accelerate just the upper bound sequences because the lower ones are
constant for all the variables. We next present the result obtained
with different methods on the variable \verb,x1, only, results
obtained with other variables are very alike.

\begin{figure}[tbp]
  \centering
 \begin{minipage}{0.48\linewidth}
\includegraphics[scale=0.25,angle=-90]{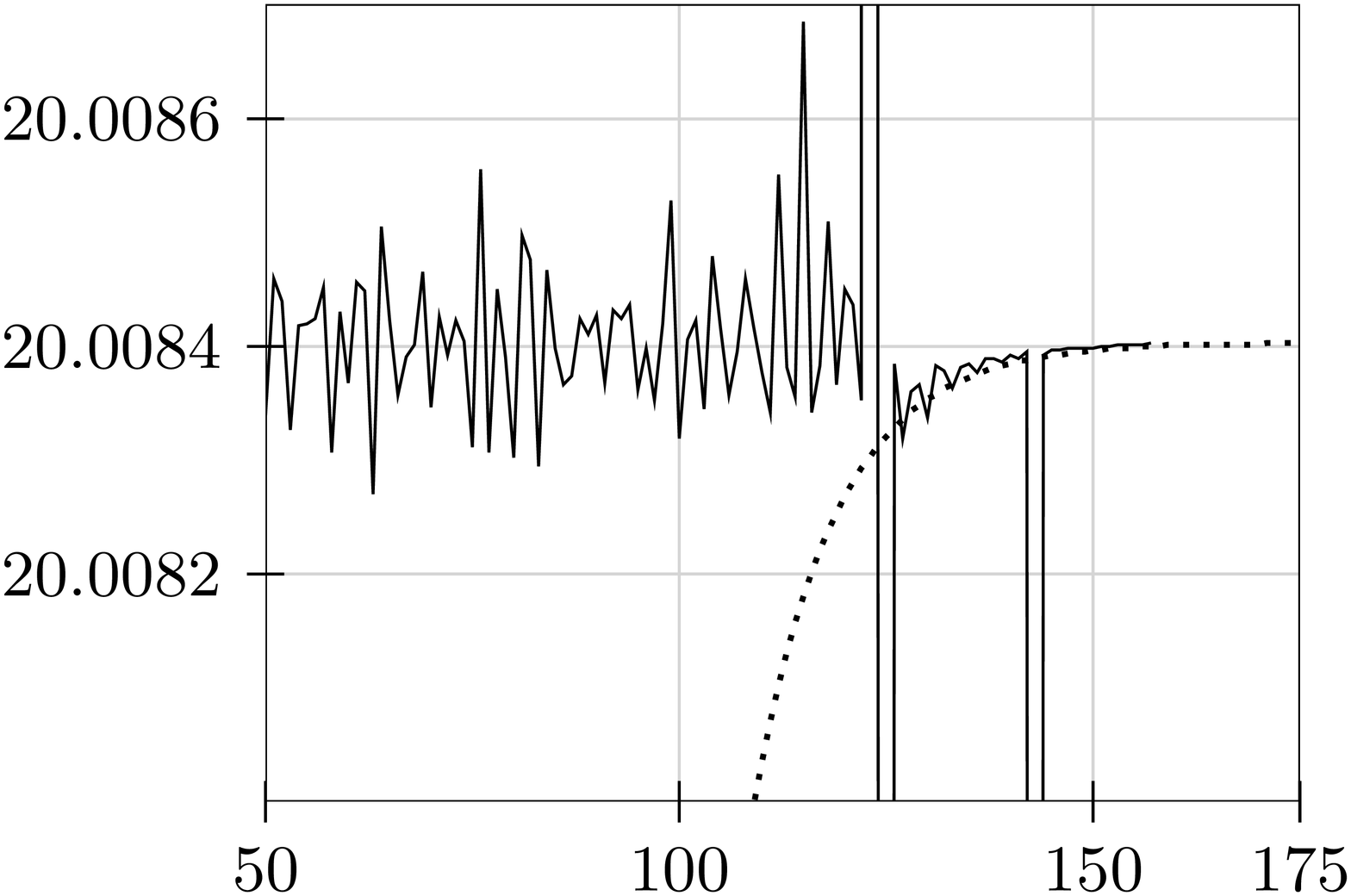}
 \end{minipage} \hfill
  \begin{minipage}{0.48\linewidth}
    \includegraphics[scale=0.25,angle=-90]{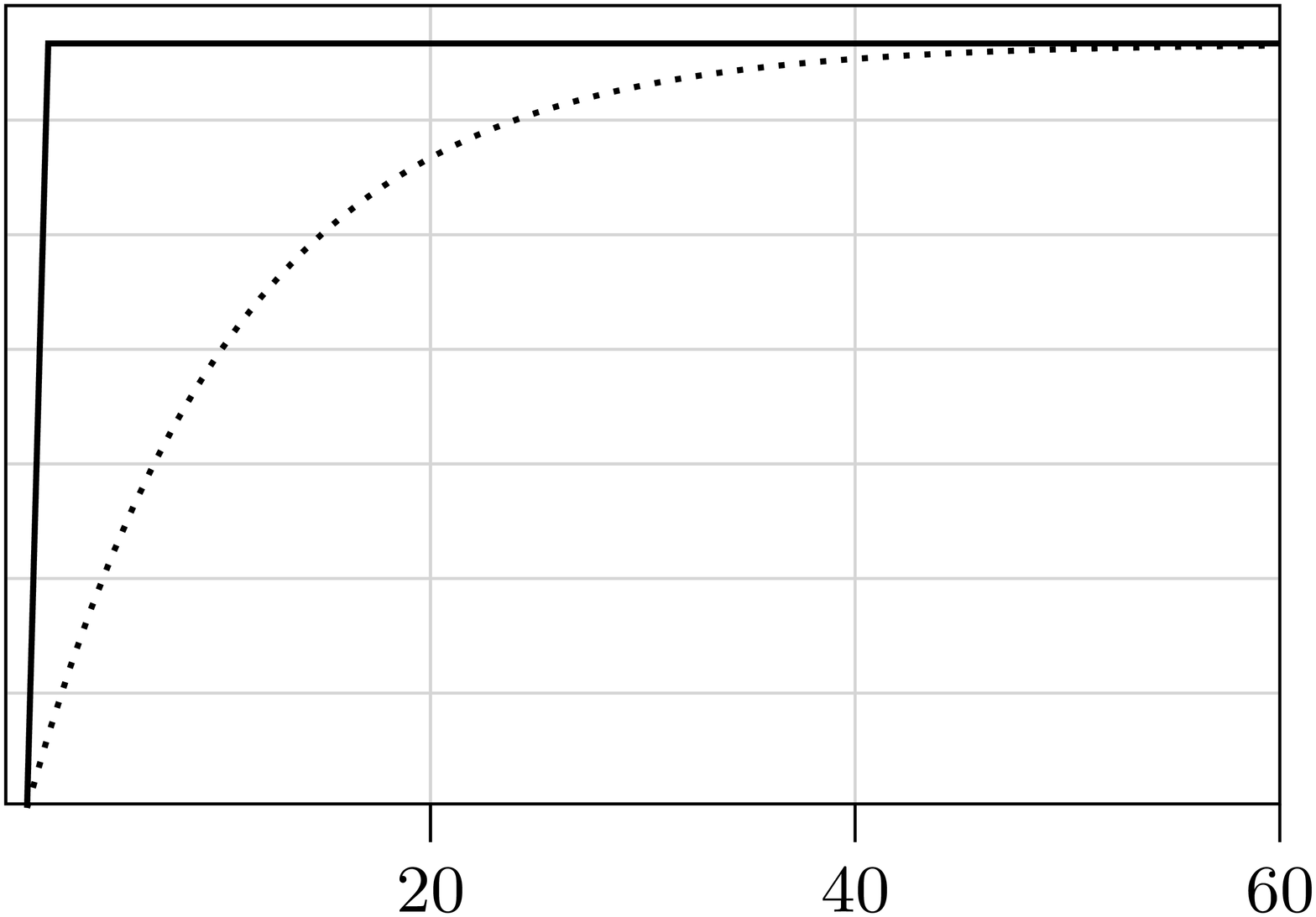}
  \end{minipage} 
 \caption{Accelerated sequences (in bold) compared with the original
   Kleene sequence (dotted). Left is the sequence obtained with Aitken
   (zooming on the numerical problems),
   right with the $\varepsilon$-algorithm (zooming on the first iterates).}
 \label{fig:butterworth-order1-accel}
\end{figure}

\null
\noindent {\bf The Aitken $\Delta^2$-method.} In
Figure~\ref{fig:prog-butterworth}, right, with Kleene iteration and
without widening, this program converges in $156$ iterations, and we
get the invariant $[0, 20.0084]$ for \verb,x1,. With the Aitken
$\Delta^2$-method, we obtain only in $3$ iterations a value very close
to $20.0084$, but problems of numerical instabilities prevent the
stabilization of the program. However the values of the accelerated
sequence stay in the interval $[20.0082, 20.0086]$ between the third
and the last iteration (see Figure~\ref{fig:butterworth-order1-accel},
left), which is a good estimate of the convergent point.

\null
\noindent {\bf The $\varepsilon$-algorithm.} In
Figure~\ref{fig:butterworth-order1-accel}, right, we notice a
important amelioration in the computation of the fixpoint, thanks to
the $\varepsilon$-algorithm. With this method, the fixpoint of the
variable \verb,x_1, is approximated with a precision of $10^{-6}$
after exactly $8$ iterations, while Kleene iteration needed $156$
steps. Remark that to obtain $8$ elements of the accelerated sequence
we need $16$ elements from the initial one. We obtain the same results
with the vector $\varepsilon$-algorithm.

\subsection{Butterworth order~$2$}
\label{sec:butterworth-order-2}

An order 2 Butterworth filter is given by the following recurrence
equation, where $\boldsymbol{x_n}$ is a two-dimensional vector, 
$\boldsymbol{x}_n=(x_1,x_2)^T$: 

{\small
  \begin{displaymath}
    \boldsymbol{x}_{n+1}=
    \begin{pmatrix}
      0.9858&-0.009929\\
      0.00929&1
    \end{pmatrix}\cdot\boldsymbol{x}_n
    +
    u\cdot 
    \begin{pmatrix}
      0.9929\\0.004965
    \end{pmatrix},\quad
    y_{n+1}=
    \begin{pmatrix}
      4.965e^{-5}\\0.01
    \end{pmatrix}\cdot\boldsymbol{x_n}+2.482e^{-5}\cdot u
  \end{displaymath}
}%

\begin{figure}[b]
  \begin{minipage}{0.65\linewidth}
    {\small
      \begin{tabular}{| c | c | c | c | c |}
        \hline
        \multirow{2}*{Variable} 
        & \multirow{2}*{Kleene} 
        & \multicolumn{3}{c|}{Vector $\varepsilon$-algorithm (\texttt{Before} + 
          \texttt{After})}
        \\
        \cline{3-5}
        & & $\delta=10^{-3}$ & $\delta=10^{-4}$ &  $\delta=10^{-5}$
        \\
        \hline
        \hline
        $x_1$ 
        & $70$ 
        & $7~(6+1)$
        & $9~(8+1)$
        & $22~(16+6)$
        \\
        \hline
        $x_2$ 
        & $83$ 
        & $26~(6+20)$
        & $23~(8+15)$
        & $17~(16+1)$
        \\
        \hline
        $y$ 
        & $83$ 
        & $26~(6+20)$
        & $23~(8+15)$
        & $19~(16+3)$
        \\
        \hline
      \end{tabular}
    }
  \end{minipage}
  \begin{minipage}{0.34\linewidth}
    \begin{small}
      \texttt{Before}: number of iterations to reach the condition on
      $\delta$. \texttt{After}: the remaining number of Kleene
      iterations to reach the invariant using the accelerated result.
    \end{small}
  \end{minipage}
  \caption{Numbers of iterations needed to reach an invariant.}
  \label{fig:results-butterworth-order2}
\end{figure}

On this program, the results obtained using the interval abstract
domain are not stable. To address this problem we have used
Fluctuat~\cite{GMP02}, a static analyzer using a specific abstract
domain based on affine arithmetic, a more accurate extension of
interval arithmetic. It returns the upper and lower bounds of each
variables. We applied the vector $\varepsilon$-algorithm on this example
with 3 different values of $\delta$ (see
Algorithm~\ref{algo:accelerated-abstract-fixpoint-computation}): this
gives Figure~\ref{fig:results-butterworth-order2}. For example, for
the variable $x_1$ and $\delta= 10^{-3}$, the over-approximation of
the fixpoint is reached after $26$ iterations ($6$ iterations before
re-injection and $20$ iterations after). Note that we obtain the same
fixpoint as with Kleene iteration. We notice that the performance of
the Algorithm~\ref{algo:accelerated-abstract-fixpoint-computation}
does not strongly depend of $\delta$. Until now, we use the
acceleration just once (unlike in
Algorithm~\ref{algo:accelerated-abstract-fixpoint-computation}), a
full implementation of it will probably reduce the number of
iterations even more.

\section{Related work}
\label{sec:related-work}


Most of the work in abstract interpretation based static analysis
concerned the definition of new abstract domains (or improvements of
existing ones), and the abstract fixpoint computation remained less
studied. Initial work from Cousot and Cousot~\cite{CC91} discussed
various methods to define widening operators. Bourdoncle~\cite{Bour93}
presented different iteration strategies that help reducing the
over-approximation introduced by widening. These methods are
complementary to our technique: as explained in
Section~\ref{sec:our-contribution}, acceleration should be done at the
same control point as the one chosen for widening, and does not
replace standard widening as the termination of the fixpoint
computation is not guaranteed. However, acceleration methods
greatly improve widening by dynamically and automatically finding good
thresholds.

Gopan and Reps in their \emph{guided static analysis}
framework~\cite{GR06,GR07} also used the idea of computing in parallel
the main iterates and a guide that shows where the iterates are going.
In their work, the precision of the fixpoint computation is increased
by computing a \emph{pilot value} that explores the state space using
a restricted version of the iteration function. Once this pilot has
stabilized, it is used to accelerate the main iterates; in a sense,
this pilot value is very similar to the value $\boldsymbol{y}_i$ of
Algorithm~\ref{algo:accelerated-abstract-fixpoint-computation}, but we
do not modify the iteration function as done in~\cite{GR07}.

Maybe the work that is the closest to ours is the use of acceleration
techniques in model checking~\cite{BFLP08}, that have recently been
applied to abstract interpretation~\cite{GH06,LS07}. In this
framework, the term acceleration is used to describe techniques that
try to predict the effect of a loop on an abstract state: the whole
loop is then replaced with just one transition that safely and
precisely approximates it. These techniques perform very well for
sufficiently simple loops working on integer variables, and gives
exact results for such cases. Again, this method is complementary to
our usage of acceleration: it \emph{statically} modifies the iteration
function by replacing simple loops with just one transition, while our
method \emph{dynamically} predicts the limit of the iterates. We
believe that our method is more general, as it can be applied to many
kinds of loops and is not restricted to a specific abstract domain
(changing the abstract domain only requires changing the $\Lambda_A$
and $\Upsilon_A$ functions).

\section{Conclusion}
\label{sec:conclusion}

We presented in this article, a technique to accelerate abstract
fixpoint computations using the numerical acceleration methods. This
technique consists in 
building numerical sequences by extracting, at every iteration,
supremum and infimum from every variable of the program. We apply to
the obtained sequences the various convergence acceleration methods,
that allows us to get closer significantly or to reach the fixpoint
more quickly than the Kleene iteration. To make sure that the fixpoint
returned by the accelerated method is indeed the fixpoint of the
abstract semantics, we re-inject it in the static analyzer. This
guarantees us the fast stop of the analyzer with a good
over-approximation of the fixpoint. The experiments made on a certain
number of examples (linear programs) show a good acceleration of the
fixpoint computation especially when we use the
$\varepsilon$-algorithm, where the number of iterations is divided by
four. Let us note that we have assumed in this article that the
sequences of iterates and the corresponding vector sequences converge
towards a finite limit. In case of diverging sequences, traditional
widening can be used as sequence transformation will not perform as well
as for converging ones. 

For now, we made the experimentation using two separate programs: one
that computes the Kleene iterates, and one that accelerates the
sequences. The
Algorithm~\ref{algo:accelerated-abstract-fixpoint-computation} is thus
still not fully implemented, its automatization is the object of our
current work. The use of the interval abstract domain allows to cover
just a small set of programs, our future work will also consist in
extending this technique to relational domains such as octagons and
polyhedra.


\bibliographystyle{plain}
\bibliography{biblio}

\end{document}